\documentclass[12pt,a4paper]{article}
\usepackage{latexsym}
\usepackage{graphicx}
\usepackage{amsfonts,amssymb}
\usepackage{cite}
\def\RE{\mbox{Re}}

\title{Understanding the sub-critical transition to turbulence in wall flows}

\author{{\large Paul Manneville}\\[0.5ex]
Hydrodynamics Laboratory, \'Ecole Polytechnique,\\
Palaiseau, F-91128, France}

\date{\footnotesize
{\sc Pramana} --- Journal of Physics, vol. 70, pp. 1009--1021}

\begin{document}
\maketitle
\sloppy

\begin{abstract}
Contrasting with free shear flows presenting velocity profiles
with inflection points which cascade to turbulence in a relatively mild
way, wall bounded flows are deprived of (inertial) instability modes
at low Reynolds numbers and become turbulent in a much wilder way,
most often marked by the coexistence of laminar and turbulent domains
at intermediate Reynolds numbers, well below the range where (viscous)
instabilities can show up \cite{DR81,SH01}.
There can even be no unstable mode at all,
as for plane Couette flow (pCf) or for Poiseuille pipe flow (Ppf) that
currently are the subject of intense research. Though the mechanisms
involved in the transition to turbulence in wall flows are now better
understood \cite{MK05}, statistical properties of the transition itself
are yet unsatisfactorily assessed. A widely accepted interpretation rests
on nontrivial solutions of the Navier--Stokes equations in the form of
unstable travelling waves \cite{KK01,Vi07,FE03,Hetal04,Ke05} and on
transient chaotic states associated to chaotic repellors \cite{EF05}.
Whether these concepts typical of the theory of
{\it temporal\/} chaos are really appropriate is yet
unclear owing to the fact that, strictly speaking, they apply when
confinement in physical space is effective while the physical systems
considered are rather extended in at least one space direction, so that
spatiotemporal behaviour cannot be ruled out in the transitional regime.
The case of pCf \cite{PD05,Betal98a,Betal98b,BC98} will be examined
in this perspective through numerical simulations of a model
with reduced cross-stream ($y$) dependence, focusing on the in-plane ($x,z$)
space dependence of a few velocity amplitudes.\cite{LM07a} In the large
aspect-ratio
limit, the transition to turbulence takes place {\it via\/}
spatiotemporal intermittency \cite{CM95} and we shall attempt to make
a connection
with the  theory of first order (thermodynamic) phase transitions,
as suggested long ago by Y. Pomeau \cite{Po86,Po98}.

\end{abstract}

\noindent{\bf Keywords}. transition to turbulence, wall flows, sub-critical
bifurcations\\
{\bf PACS Nos 47.27Cn, 47.27Lx, 47.20Ky}

\section{Introduction}

The transition to turbulence is an important long standing problem
owing to the marked difference between transport properties of laminar
and turbulent flows but the process can follow different scenarios
depending on the physical situation under consideration.

In {\it closed\/} flows,
besides the instability mechanisms, e.g. Rayleigh--B\'enard or
B\'enard--Marangoni for convection, Taylor--Couette for centrifugal
flows \cite{DR81}, lateral effects play an important role. One can
distinguish
{\it confined\/} systems  from {\it extended\/} ones. In confined
systems, all the dimensions of the experimental cell are of the order
of the length over which the mechanism operates,
{\it bifurcation theory\/} applies to a limited set of
{\it central modes\/} governed by {\it normal forms\/} obtained
through adiabatic elimination of enslaved modes. The classical
scenarios of transition to {\it temporal chaos\/} follow. In extended
systems, at least one of the transverse dimensions is much larger
than that in the direction selected by the instability mechanism
and, while the latter still generates cells at a local scale, large
scale modulations well described within the {\it envelope formalism\/}
can degenerate into {\it spatio-temporal chaos\/} \cite{Ma90}.

\begin{figure}
\begin{center}
\includegraphics[height=0.3\textwidth,clip]{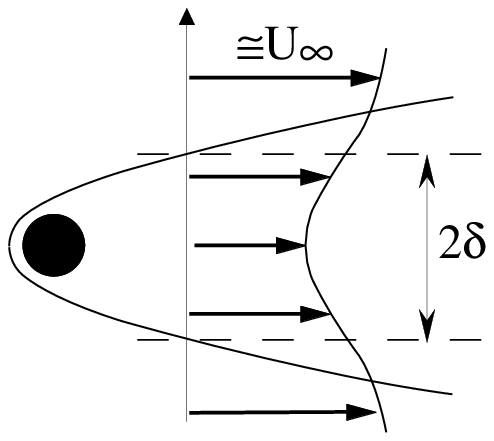}
\hskip2ex
\includegraphics[height=0.3\textwidth,clip]{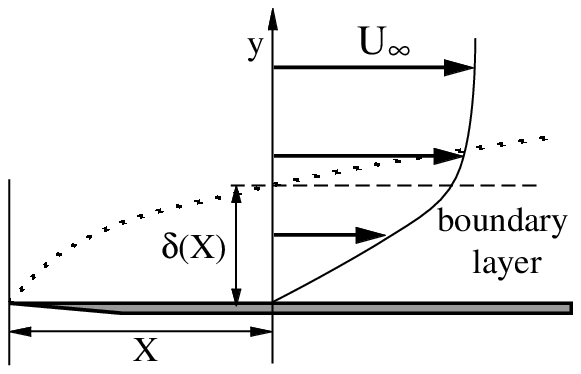}
\end{center}
\caption{The wake of a blunt obstacle (left) and the boundary layer
  flow (right) are examples of inflectional and non-inflectional base
  flow profiles,   respectively.\label{f1}}
\end{figure}

In {\it open\/} flows, the situation is more complex and
less well understood. The natural control
parameter is the Reynolds number $\RE=\Delta U\Delta \ell/\nu$, where
$\Delta\ell$ is the typical length scale, $\Delta U/\Delta \ell$ the
typical shear in the system, and $\nu$ the kinematic viscosity
measuring dissipation effects.
At the {\it linear stage\/} \cite{DR81,SH01}, standard stability analysis
helps one classify base flow profiles into  {\it inflectional\/}
and {\it non-inflectional\/} (fig.~\ref{f1}).
On general grounds, inflectional profiles become unstable and then
turbulent through linear instability mechanisms of inertial origin
(Kelvin--Helmholtz mechanism) at rather low $\RE$ and {\it via\/}
cascading scenarios with mild super-critical flavour, i.e. with the
bifurca{\it ted\/} state staying in some sense close to the
bifurca{\it ting\/} one.
On the contrary non-inflectional flow profiles experience no
instability at low $\RE$ but can possibly become unstable against
subtle linear viscous mechanisms (Tollmien--Schlichting waves) at large
$\RE\ge\RE_{\rm TS}$ only.

An essential assumption of linear stability analysis is the
mathematically infinitesimal character of the perturbations. Relaxing
this assumption one finds that, when $\RE_{\rm TS}$ is
large, the flow can depart from its laminar
profile at values of $\RE<\RE_{\rm TS}$. {\it Conditional stability}
is thus expected to be the rule for non-inflectional base flow
profiles.

\begin{figure}
\begin{center}
\includegraphics[height=0.25\textwidth,clip]{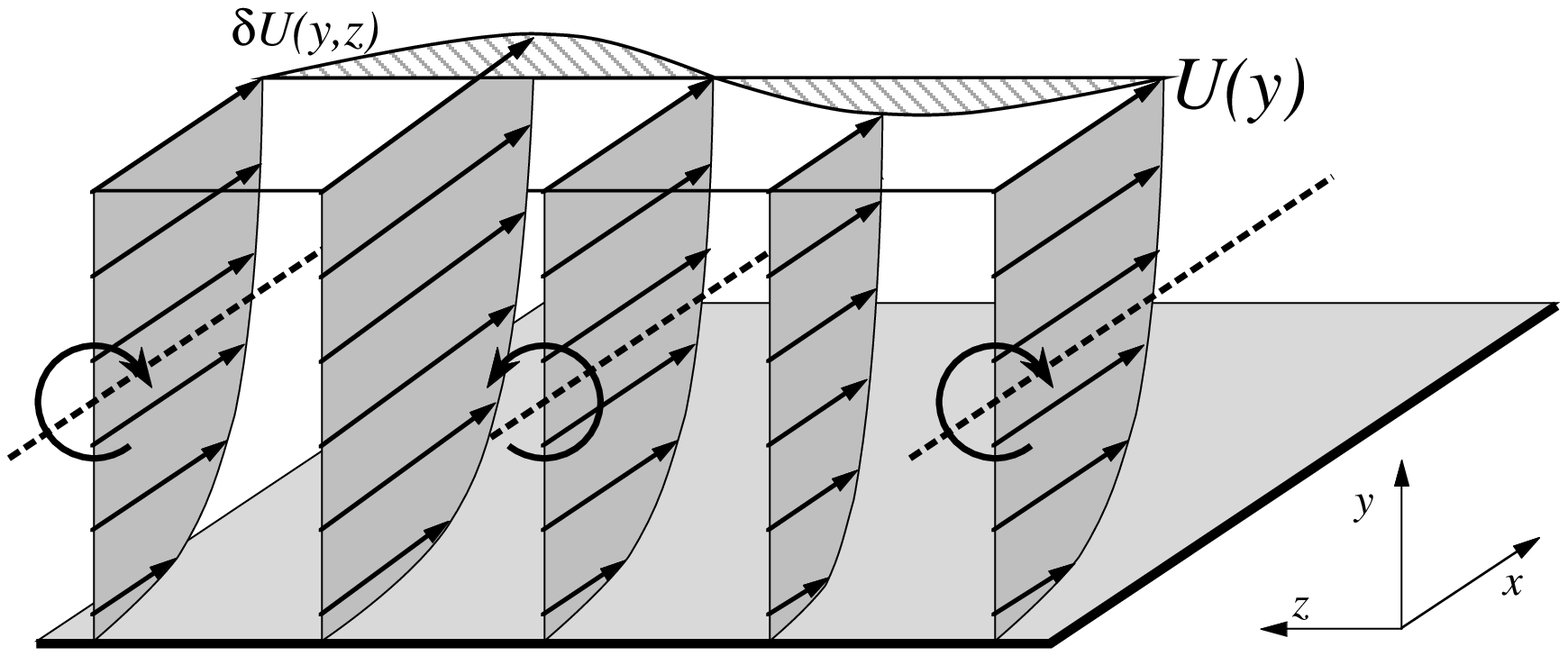}
\hskip1.5em
\includegraphics[height=0.25\textwidth,clip]{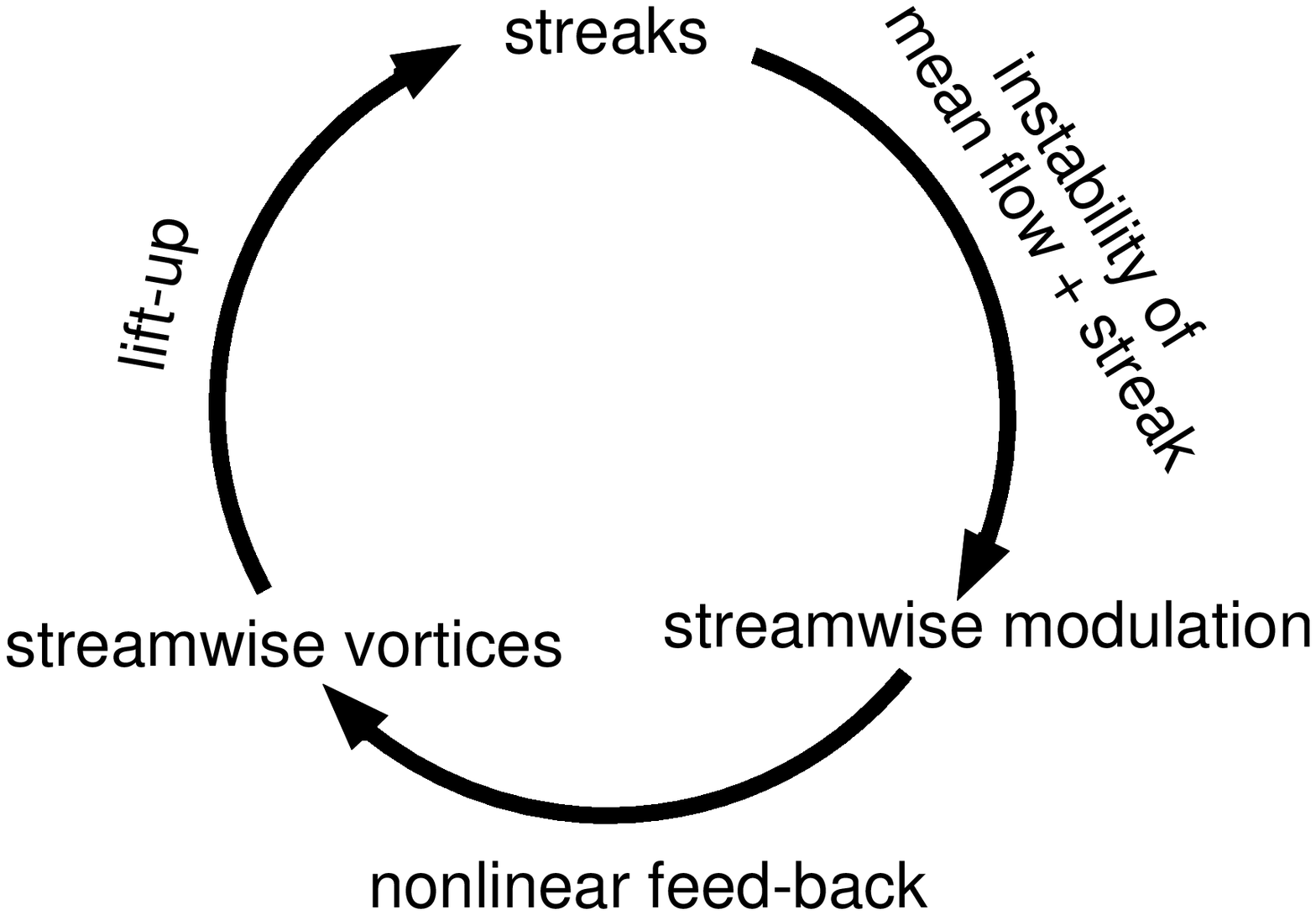}
\end{center}
\vspace{1ex}

\caption{Left: Lift-up mechanism by which streamwise vortices induce
  alternatively slow and fast streaks. Right: Self-sustaining process in
  wall flows.\label{f2}}
\end{figure}

The physical role of advection in open flows is worth considering 
(fig.~\ref{f2}, left). Assuming a perturbation in the form of
streamwise vortices (i.e. with axes aligned along the flow
direction), it is immediately seen to induce
flow corrections, called {\it streaks\/}, that are alternatively
slowed down and accelerated with respect to the base flow. This mechanism,
called {\it lift-up\/}, leads to transient perturbation energy growth even
in linearly stable flows,%
\footnote{Mathematically, transient energy growth is linked to the {\it
 non-normality} of the linear stability operator, i.e. the fact that it
does not commute with its adjoint so that eigenvectors are not
orthogonal~\cite{SH01}.}
thus plays a role in the direct
transition to turbulence, and is an indisputable ingredient of
the sustainment of turbulence in wall flows (fig.~\ref{f2}, right),
see \cite{MK05,WW05}.

At the nonlinear stage, the
{\it by-pass\/} scenario usually involves the nucleation and growth of
{\it turbulent spots\/} embedded in linearly stable laminar flow. An
example of turbulent domain (mature spot) immersed into laminar 
plane Couette flow (pCf) is given in figure~\ref{f3} (left).
Laminar-turbulent coexistence can be observed in other wall
flows such as plane Poiseuille channel flow or the boundary layer
flow. The shape and relative speed of spots depend on the case
studied, see figure~1 of \cite{PD05} for illustrations.
Owing to the absence of overall advection and of TS instability
mode ($\RE_{\rm TS}=\infty$),
plane Couette flow seems to be the simplest possible case to study.
Poiseuille pipe flow (Ppf) is another classical example of flow that is
always linearly stable \cite{DR81} but becomes turbulent due to nonlinear
perturbations. There, the coexistence of laminar and turbulent flows
take the form of {\it turbulent puffs\/} becoming
{\it turbulent slugs\/} at larger $\RE$ \cite{WC73}.
These systems have been the
subject of intense study recently. Here we consider the case of pCf 
and keep in mind Ppf results \cite{Hetal06,PM06,WK07} for comparison. 

\section{Phenomenology of plane Couette flow}

Plane Couette flow is the flow ideally obtained by shearing a fluid between
two infinite parallel plates at a distance $2h$ moving in opposite directions
at speeds $\pm U$, which defines the streamwise direction $x$, $y$ and
$z$ being the wall-normal and spanwise directions, respectively. The
laminar profile is just $U_{\rm b}(y)=Uy/h$. It
is known \cite{DR81} to stay linearly stable for all values of the Reynolds
number $\RE=Uh/\nu$ ($\nu$ is the kinematic velocity) but, of course,
to become turbulent for large enough $\RE$. 
Here we summarise experimental results obtained by the Saclay group
\cite{Detal92,DD94,Betal98a,Betal98b,BC98,Petal02,PD05}.

Basically, three kinds of experiments were performed:
i$_1$) {\it spot triggering\/}, a tiny impulsive jet of controlled
intensity is sent through the flow \cite{Detal92,DD94,BC98};
\begin{figure}
\includegraphics[height=0.2\textwidth,clip]{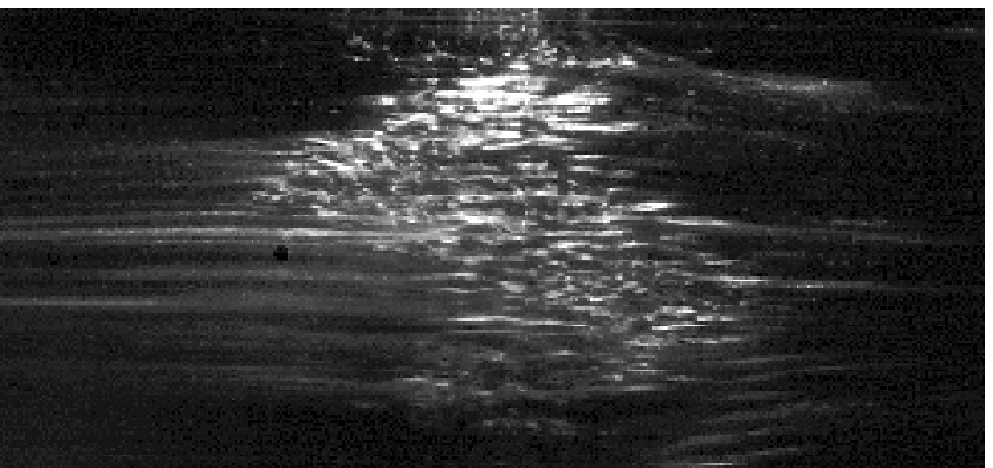}
\hfill
\includegraphics[height=0.2\textwidth,clip]{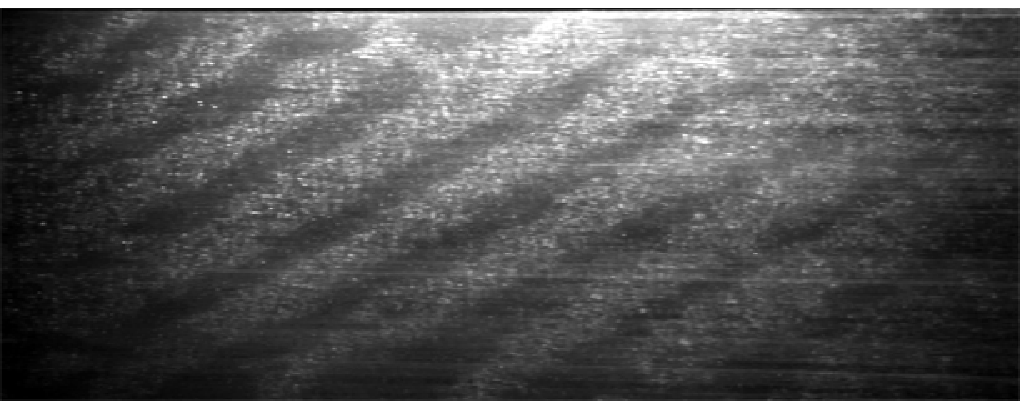}
\caption{Left: spot in plane Couette flow at a late stage of its
  evolution; the streamwise direction is horizontal and the streaks
  are clearly visible; the hole through which the triggering jet is
  shot appears as a dark dot on the left of the picture's centre
  (courtesy S.~Bottin). Right: turbulent stripes at the upper end of
  the transitional regime (courtesy A.~Prigent). In units of the half-gap
  $h$, the size of the set-up was $380h\times2h\times70h$ for the left
  snapshot and $770h\times2h\times340h$  for the right one.\label{f3}}
\end{figure}
i$_2$) {\it quench experiments\/}, turbulent flow is prepared at some
initial high value of $\RE_{\rm i}$ and $\RE$ is 
suddenly decreased to some final value $\RE_{\rm f}$
\cite{Betal98b,BC98};
and ii) variable-strength {\it permanent modifications\/} to the base
flow \cite{Betal98a}. Experiments of kind (ii) which approach pure
Couette flow by a continuation method confirm the results of
experiments of kind (i) which are basically initial value problems.
The bifurcation diagram given in figure~\ref{f4} summarises the
outcome of kind-(i) experiments:\\
-- For $\RE <\RE_{\rm u}\simeq280$, the laminar
profile is rapidly recovered whatever the intensity of the
perturbation brought to the flow.\\
-- For $\RE_{\rm u}<\RE\lesssim\RE_{\rm g}=325$, turbulence is only transient
but as $\RE_{\rm g}$ is approached from below, the lifetime of
turbulence increases. For $\RE_{\rm g}\lesssim\RE\lesssim360$
turbulence takes the form of irregular large spots (fig.~\ref{f3},
left).%
\footnote{Early reports on experimental or numerical spot generation
  \cite{TA92,Detal92,LJ91} mention values of $\RE$ in the range
  $360$--$370$. Large long-lived turbulent patches observed for
  $\RE\lesssim360$ were rather obtained in quench-type experiments.}
The actual status of $\RE_{\rm g}$ is discussed in more detail below.\\
-- For  $360\lesssim\RE\lesssim\RE_{\rm t}=415$ the spots merge to
form oblique stripes (fig.~\ref{f3}, right). These stripes are characterised
by a regular modulation of the turbulence intensity which dies out
when $\RE_{\rm t}$ is approached from below, leaving one with a regime
of featureless turbulence \cite{Petal02,PD05}.
\begin{figure}
\begin{center}
\includegraphics[width=0.98\textwidth,clip]{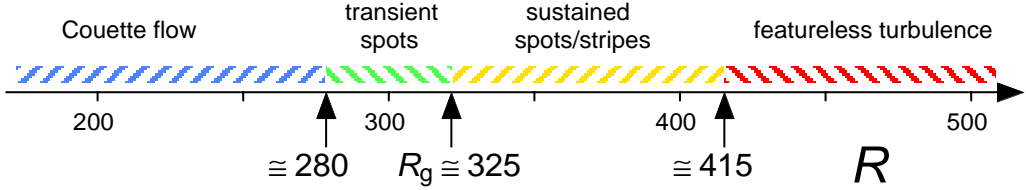}
\end{center}
\caption{Bifurcation of the plane Couette flow as obtained by the
  Saclay group.\label{f4}}

\end{figure}

\begin{figure}
\centerline{
\includegraphics[width=0.49\textwidth,clip]{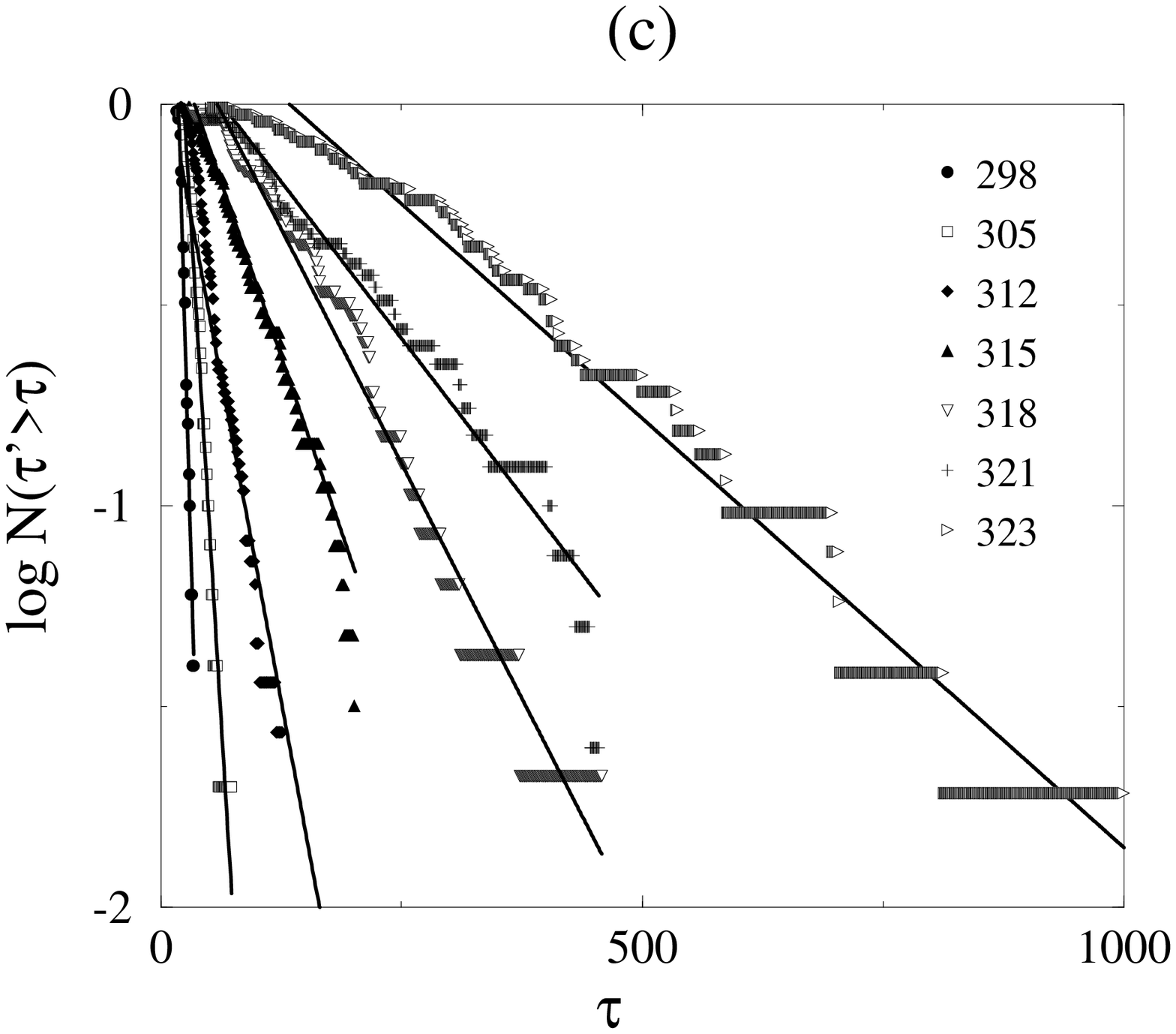}
\hfill
\includegraphics[width=0.49\textwidth,clip]{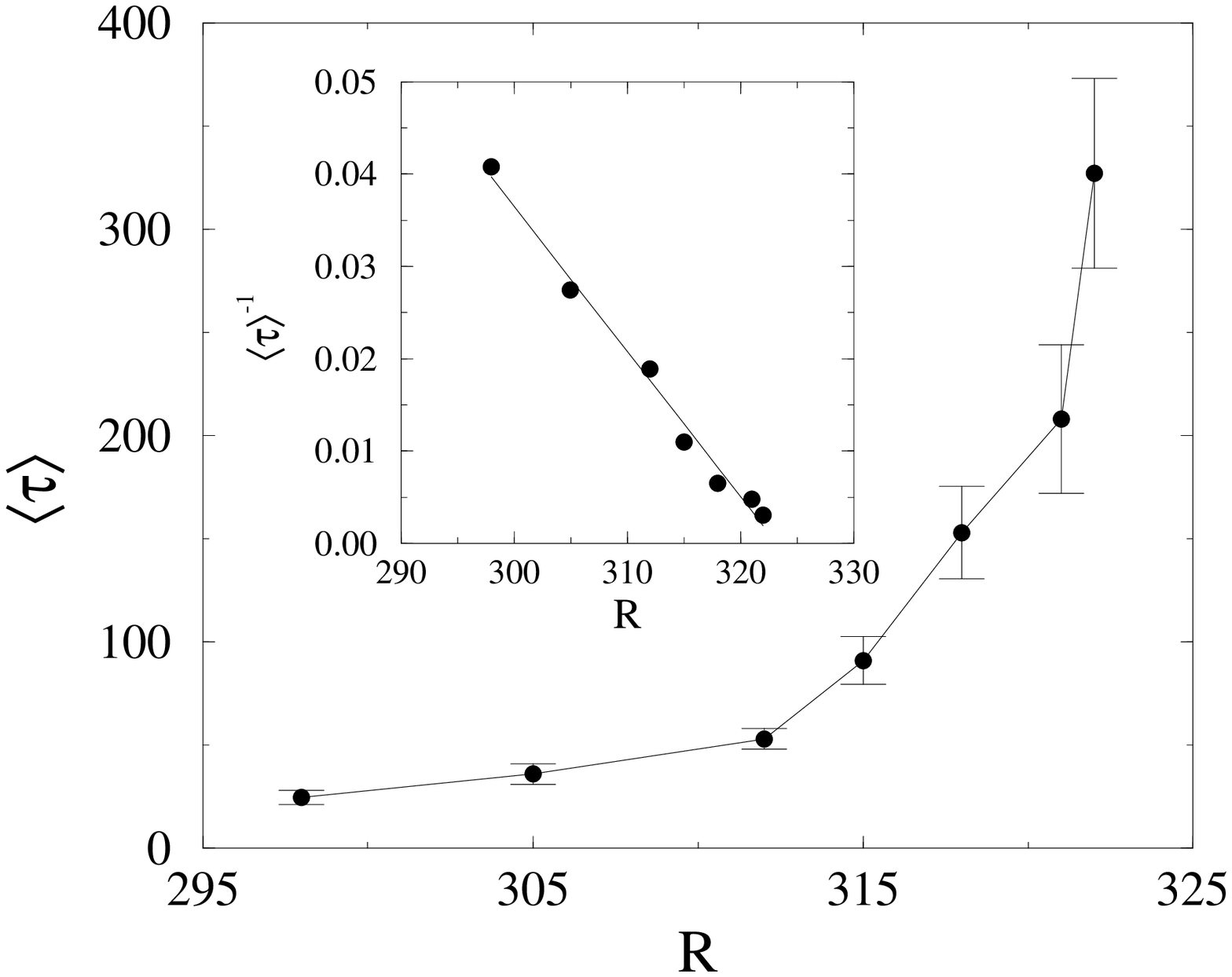}}
\caption{Left: distributions of transient lifetimes for a series of
  Reynolds numbers (quench experiments). Right: characteristic times
  extracted from these distributions as functions of $\RE$.
Courtesy S.Bottin.\label{f5}}
\end{figure}

Here we concentrate our attention on the lowest part of the
transitional regime, i.e. $\RE\lesssim360$ where turbulent patches are
not yet organised in oblique stripes. In the diagram, $\RE_{\rm g}=325$
presents itself as the global stability threshold (`g' for `global'),
i.e. the value of 
the control parameter below which the final state is always the
laminar profile, whatever the amplitude of the initial triggering
perturbation. Below $\RE_{\rm g}$ turbulence is therefore not
sustained. The lifetimes of transients were found to be distributed
according to exponentially decreasing laws in the form
$\mathcal{N}(\tau'>\tau)\propto\exp(-\tau/\langle\tau\rangle)$, see
figure~\ref{f5} (left). The early proposal, made by Bottin
and Chat\'e \cite{BC98}, that the characteristic times
$\langle\tau\rangle$ of the distributions were diverging as
$1/(\RE_{\rm g}-\RE)$, was extracted
from the results by simply taking the mean of the
transients' lifetimes at given $\RE$ (fig.~\ref{f5}, right). However,
these results were somewhat noisy and $\RE_{\rm g}$ was not approached
sufficiently closely to make the conclusion decisive. The methodology
giving $\langle\tau\rangle$ was further criticised by Hof {\it et al.}
\cite{Hetal06} who rather suggested the absence of a ``critical'' point%
\footnote{In the Landau theory of second order phase transitions
  ($\equiv$ forward pitchfork bifurcation)
  critical slowing down corresponds to a divergence of the
  relaxation time as $\langle\tau\rangle\sim(\RE_{\rm c}-\RE)^{-1}$.
  By abuse of language, power law divergence
  will be called ``critical'' in the following.}
and an exponential increase of the characteristic time with
$\RE$. This new proposal, made in close 
correspondence with their findings in the related problem of
transitional Ppf, fully agrees with our reanalysis of the original
data by fitting the terminal parts of the distributions in
figure~\ref{f5}~(left) against straight lines.
However the extrapolation of the exponential variation of the characteristic
time with  $\RE$ to much larger values
may not be justified owing to the limited range of values studied.

According to Hof {\it et al.}, transient behaviour with exponential
distribution of lifetimes is associated to {\it homoclinic tangles\/}
\cite{EF05},
themselves resulting from the presence of nontrivial unstable periodic
orbits in a low dimensional dynamical systems perspective. Such
solutions are known to exist both in Ppf \cite{FE03,Hetal04,Ke05} and in 
pCf \cite{KK01,Vi07} and are furtively observed in
experiments \cite{Hetal04}. Since Poincar\'e's work, the way a transverse
intersection of stable and unstable manifolds of a limit cycle
generates an uncountable infinity of intersections is well
understood, as well as how complexity enters, through the symbolic
dynamics attached to the modelling of that tangle in terms of Smale's
horseshoe. The exponential distribution of lifetimes of transient
chaotic trajectories around the corresponding chaotic repellor then
follows but, unfortunately, this nice application of the theory
of low dimensional dynamical systems cannot predict how the slopes of
the distributions vary with $\RE$. In this respect, the case of
pPf is not completely settled, some results displaying exponential
behaviour \cite{Hetal06}, others a ``critical'' behaviour
$\propto 1/(\RE_{\rm g} -\RE)^\alpha$ for some global threshold
$\RE_{\rm g}$  with $\alpha\sim1$ \cite{PM06,WK07}. A
first reason explaining the observed discrepancies could be the
experimental conditions since some experiments were performed at 
constant driving pressure gradient \cite{Hetal06} and others at
constant mass flux \cite{PM06,WK07}. As suggested by our results
on plane Couette flow, another plausible possibility
is linked to the fact that the physical system is not confined,%
\footnote{In most numerical approaches, the application of periodic
  boundary conditions at a short distance in the streamwise direction,
  typically 5 diameters, transforms the initial non-confined problem
  into a strongly confined one, hence a finite dimensional dynamics.}
but quasi-1D in physical space, i.e. confined in the radial
direction, discretised in the azimuthal direction, but extended in
the streamwise direction. Recent results from R.R. Kerswell and coll.
(private communication) seem to point in that direction.

Taking for granted that the reduction of the dynamics to a 0D problem
in physical space leaves aside interesting questions about the
transitional regime in pCf, even more than in Ppf, we now study
its dynamics in a quasi-2D spatiotemporal context, i.e.
depending on space in the streamwise ($x$) and spanwise ($z$) directions,
while keeping confinement conditions in the cross-flow direction ($y$). 

\section{Modelling transitional plane Couette flow}

In contrast with most numerical studies restricting the system size
by placing in-plane periodic boundary conditions at small distances
(the size of the so-called {\it minimal flow unit\/} is typically
$4\pi h\times2h\times2\pi h$ where $h$ is the half-gap \cite{Letal07})
and implicitly analysing
the results in a finite-dimensional dynamical systems framework,
experiments reported above were performed in domains at least as large
as $380h\times2h\times70h$.
Direct numerical simulations of the full Navier--Stokes
equations in such domains are indeed not yet feasible with accuracy%
\footnote{A quasi-1D fully resolved approach was followed by Barkley and
  Tuckerman \cite{BT07} who took an oblique reference direction
  to study turbulent stripes observed in the upper transitional range
  pictured in  fig.~\ref{f3} (right).}
so that we have found it advisable to develop a spatiotemporal model.
Instead of a set of differential equations governing scalar amplitudes
(Lorenz-like model), the resulting model was a set of partial
differential equations for a few fields (Swift--Hohenberg-like model).
It was obtained  from primitive equations (3D) through a systematic
{\it weighted residual approach\/}, the Galerkin method, which uses
a basis that fulfils the no-slip boundary conditions and projects the
residuals on the same basis \cite{Fi72}. Here a simple polynomial
basis was chosen and the expansion was truncated at lowest nontrivial
order thus freezing most of the cross-flow space dependence, while
leaving the in-plane dependence free. This was justified by the fact
that, in the lowest part of the transitional regime, experimental and
numerical evidence suggests the existence of coherent patterns
occupying the full gap $2h$ with limited cross-stream structure
\cite{Betal98a,KNN05}.  This restriction could of course be overcome
but at a price of heavier analytical and numerical computations that
we are not ready to pay for, since we mostly look for qualitative
hints and not for quantitative agreement.

Due to the way it is obtained, the model {\it a priori\/} displays
the problem's most relevant general properties such as non-normal linear
terms accounting for lift-up mechanism, linear viscous damping, nonlinear
advection terms preserving perturbation kinetic energy, and linear
stability of the base profile for all $\RE$.
Numerical simulations {\it a posteriori\/} show that it also shares
many features of the complete physical system. In particular
the statistics of homogeneous turbulent state no longer depends on the
size of the simulation domain as soon as it is large enough (extensivity
property) while the  {\it turbulent$\>\to\>$laminar\/}
transition is indeed discontinuous with exponentially distributed
transient lifetimes \cite{LM07a}. 

Using this model, our main objective has been to contribute to
``critical/exponential'' controversy in the pCf case (with possible
extrapolation to the Ppf case) and more particularly to point  
out the possible role of size effects.

\subsection{Sub-criticality in the $32h\times2h\times32h$ system}

In a first instance a $32h\times2h\times32h$ system has been considered,
which is of moderate size when compared to the size of coherent
structures $\sim12h\times2h\times6h$ \cite{WW05}.
The state of the system was determined from its mean turbulent
energy contents. The global stability threshold, with all the ambiguities
the term covers, was determined by a combination of quench
experiments, where $\RE$ is abruptly decreased from $\RE_{\rm i}=200$ for
which the flow is uniformly turbulent
to variable $\RE_{\rm f}\ll\RE_{\rm i}$, and annealing experiments where 
$\RE$ was quasi-adiabatically decreased.
As a result, turbulence seemed sustained for $\RE$ greater than
$\approx175$ but
definitely transient  for $\RE<175$: the time series of the mean turbulent
energy presented well distinct long plateaus before unpredictable sudden
decay happened. Furthermore, the distributions of the transients' lifetimes
were clearly exponential and closely resembled the experimental ones
depicted in figure~\ref{f5} (left). The difference between the global
threshold in the model $\RE_{\rm g}\approx175$ and in the laboratory
$\RE_{\rm g}\approx325$ could be attributed to the under-estimation of
viscous dissipation and energy transfer towards small cross-stream
scales due to truncation. In spite of this lowering of the transitional
regime by an empirical factor of about two, most qualitative aspects of
nonlinear processes seemed preserved, e.g. those linked to the
development of turbulent spots \cite{LM07b}.

\begin{figure}
\includegraphics[width=0.49\textwidth,clip]{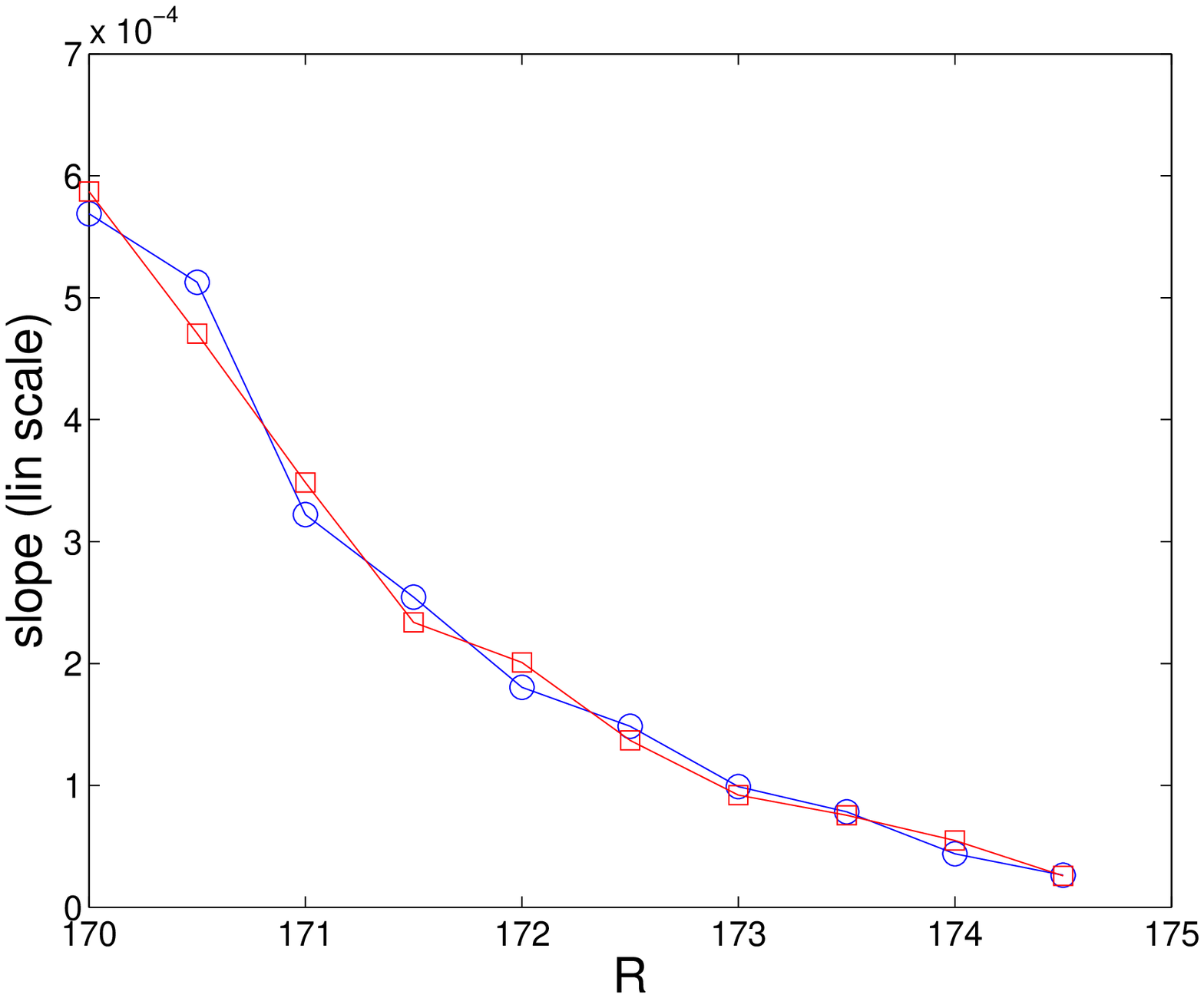}
\hfill
\includegraphics[width=0.49\textwidth,clip]{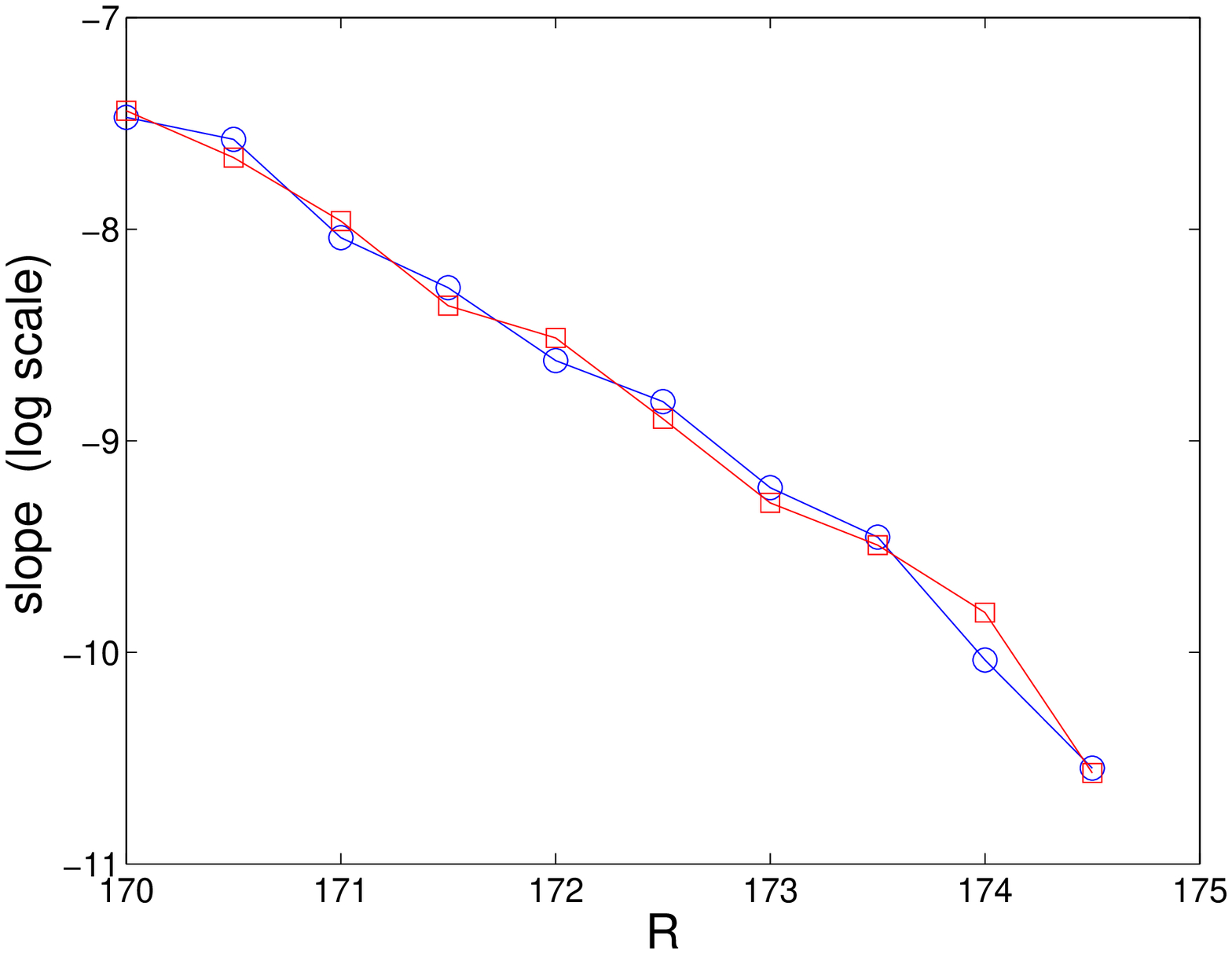}
\caption{Decay rate of lifetime distributions as functions of $\RE$ in
  lin-lin scale (left) and in lin-log scale (right).\label{f8}}
\end{figure}

The variation with $\RE$ of the decay rate of the lifetime histograms
is given in figure~\ref{f8}, which shows well aligned
points in lin-log scale, hence exponential behaviour, except for $\RE=174$
and $174.5$ which are mis-aligned. It turns out that this mis-alignment
could not be explained by statistical errors, which suggests a crossover
to critical behaviour very close to $\RE=175$, possibly linked to
size effects. 
Improving over this result was however beyond reach of numerical
means and the visualisation of the velocity fields during decay did not
help us discriminate between temporal and spatio-temporal behaviour. 

In view of the typical size of streak segments mentioned earlier,
temporal behaviour remained plausible in the $32h\times2h\times32h$
system. Expecting that this will no longer be the case when
confinement effects are made weaker, we considered a much larger
system of size $256h\times2h\times128h$. 

\subsection{Sub-criticality in the $256h\times2h\times128h$ system}

\begin{figure}
\includegraphics[width=0.55\textwidth,clip]{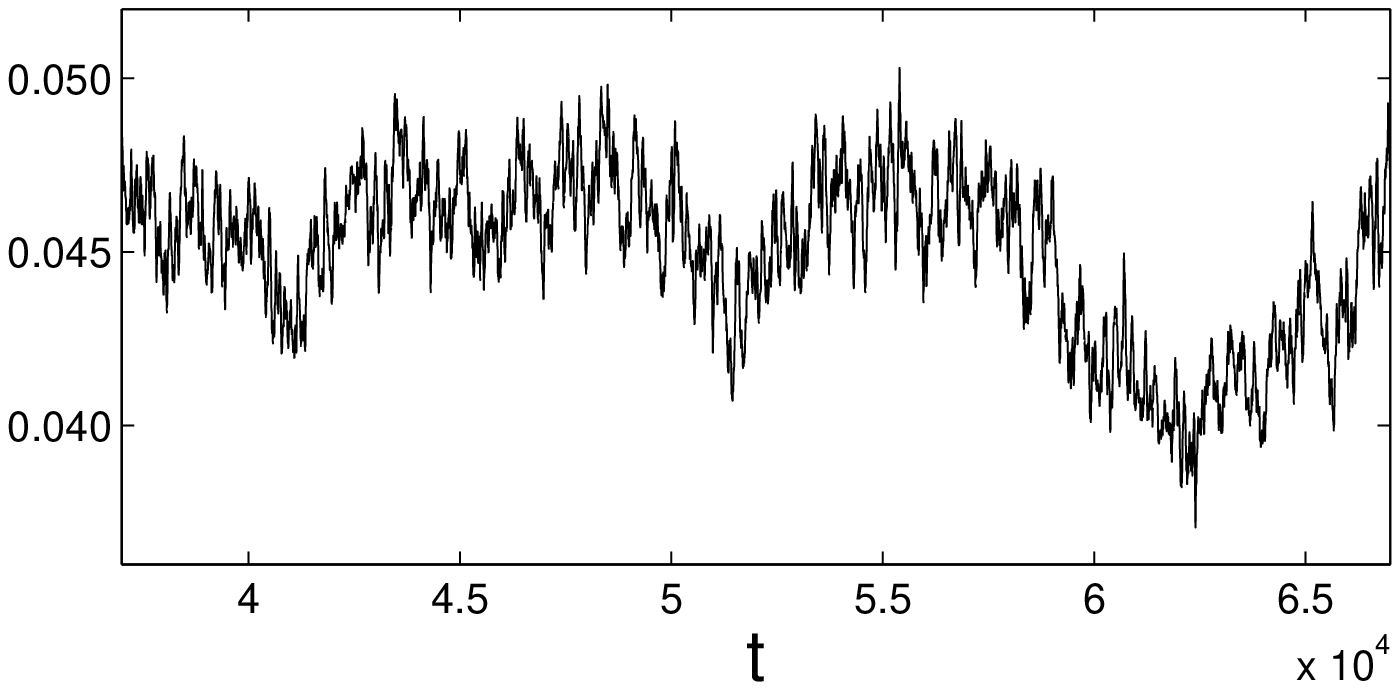}
\hfill
\includegraphics[width=0.43\textwidth,clip]{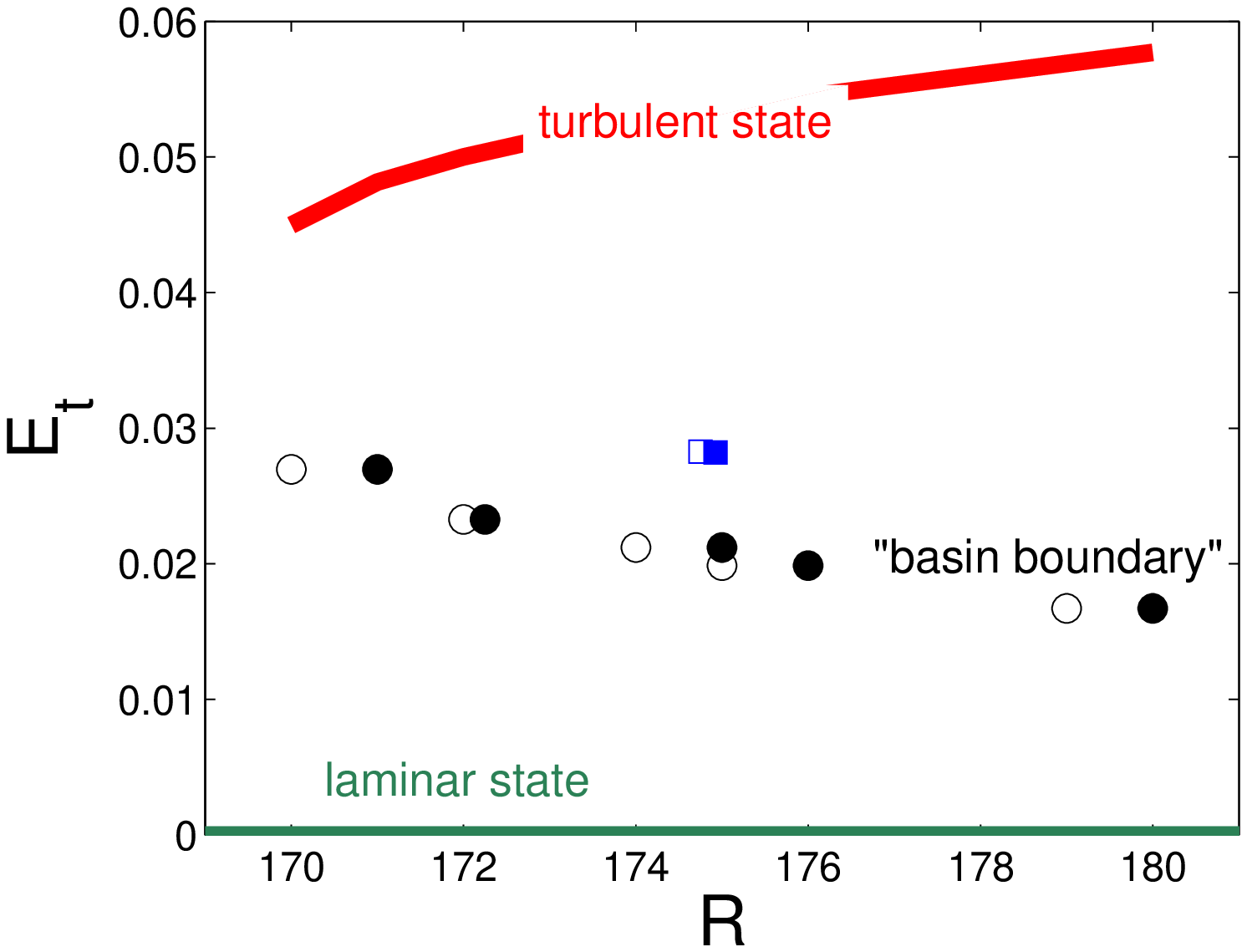}
\caption{Left: Time series of the mean turbulent energy for the
$256h\times2h\times128h$ system at $\RE=170$. Right: Bifurcation
diagram as obtained from the model in same system
(full line: average mean turbulent energy during the adiabatic decrease of
$\RE$; open and filled dots: bracketing of the laminar/turbulent boundary
for homogeneous low-level turbulent initial conditions; open and
filled square: bracketing of the frontier at a single point for a
strongly inhomogeneous initial state.\label{f10}}
\end{figure}

Annealing experiments ($\RE$ decreasing quasi-adiabatically) first
showed that turbulence could be maintained for very long times well
below $\RE_{\rm g}\simeq175$ without any sign of decay. The time
series of the mean turbulent energy shown in figure~\ref{f10} (left) was
obtained in this way for $\RE=170$. In contrast, relaxation at the end of a
very long but regular and monotonic transient was observed for $\RE=169$
without any trace of plateau indicating that turbulence could be metastable.
So, at least down to $\RE=170$ the turbulent state thus seems to be a local
attractor and the continuous line in figure~\ref{f10} (right) indicates
the variation of the corresponding mean turbulent energy as a function
of $\RE$.

Visualisations of the turbulent pattern at $\RE=170$ showed that,
when the mean turbulent energy level was high, the system was in a state of
fine-grained mixture of turbulent and laminar patches ({\it spatiotemporal
intermittency\/} \cite{CM95}), whereas during the excursions toward
comparatively low values, large laminar domains were present for relatively
long lapses of time. An explanation to the exponential behaviour of the
lifetime distributions, alternative to the accepted one in terms of chaotic
saddles, could then be obtained by considering the large
$256h\times2h\times128h$ system as an assembly of 32 smaller
$32h\times2h\times32h$ sub-systems and comparing the typical dynamics
of the sub-systems to that of the $32h\times2h\times32h$ system \cite{Maxx}.
Returning to the smaller system and comparing its time-series histograms
at $\RE=200$ to those at $\RE=175$, it could be seen that an
exponential tail appeared  at low energy when $\RE$ was decreased, so
that, when $\RE$ was further decreased, excursions toward smaller and
smaller energies were more frequent, forcing the decay of a given
transient when some limiting energy,
$E_{\rm t}^{\rm lim}(32\times32)\sim0.025$, was reached.
Setting this observation in the context of the sub-systems' dynamics,
one could then appeal to Pomeau's analogy between sub-critical
bifurcations in extended systems and {\it first-order (thermodynamic)
phase transitions\/} \cite{Po86} and the related theory of
nucleation: The advent of a sizable laminar domain in the large 
system implies an excursion of the mean energy towards smaller values
(e.g. $t~\sim41000,\, 51000,\, 62500$ in fig.~\ref{f10}, left), though the
system apparently remains in the sustained turbulent regime. 
Such excursions correspond to the breakdown of turbulence over
regions already larger than the size of a sub-domain, which is also
the size of the smaller system. Accordingly, while turbulence is
relatively short-lived in the smaller system (since the occurrence of
such a fluctuation would have led the turbulent regime to its end), it
can be long-lived in the larger system because a wide region that fell
laminar can become turbulent again by contamination from its
surroundings. This observation opens the way to the understanding
of the whole `turbulent~$\to$~laminar' transition in terms of
directed percolation and statistical estimates that come with it
\cite{Po98}.

The interpretation of the exponential decay of lifetime
distributions in the smaller system thus rests on the idea that
spatiotemporal fluctuations result in a random process exploring
the low energy exponential tails of the mean turbulent
energy histograms appearing when $\RE$ is small enough. While such a
tail was indeed unobservable at $\RE=200$, it was already sizable at
$\RE=175$, and quite substantial for $\RE=170$ provided
that the histogram was determined under the condition that the 
system is still in the chaotic transient state, i.e. as long as
$E_{\rm t}>E_{\rm t}^{\rm lim}(32\times32)$. The argument was closed
by saying that the uniformly random exploration of the low-energy
exponential histogram tails converted itself into exponentially
decreasing lifetime histogram tails.%
\footnote{This observation suggests that turbulence is in fact still
  not sustained at $\RE=175$.}
Unfortunately, like the conventional view, this new interpretation does not
predict how the lin-log slopes vary since it does not tell us how
the tail's importance changes with $\RE$ though the trend can be easily
guessed. Invoking a spatiotemporal origin to the shape of the lifetime
distribution could however help us understand the possible presence of a
cross-over from exponential to critical variation.

Evidence that the turbulent state for $\RE>170$ is a
local attractor comes from the attempt to determine the frontier of its
attraction basin as seen from the {\it laminar $\to$ turbulent\/}
viewpoint. Open and filled dots in figure~\ref{f10} (right)
bracket the ``line''
separating random initial conditions with given initial mean turbulent
energy obtained by attenuating the same homogeneous turbulent solution
with variable factors. Above the line, the system evolves toward the
turbulent state whereas it relaxes below.
However the frontier appears to be strongly dependent on the shape of
the initial condition and this dependence is better understood in physical
space than in phase space.
For example the open and filled squares in figure~\ref{f10}
(right) bracket the frontier corresponding to another kind of initial
condition displaying wide laminar domains competing with the
spatiotemporally intermittent state alluded to above. Still another
frontier would be obtained for transverse turbulent stripes (parallel to
the $z$ axis, not related to stripes in fig.~\ref{f3}) that are found
to invade the system only when 
$\RE>\RE^{\perp}$, where  $\RE^{\perp}$ is the threshold value for
spanwise turbulent stripes and happens to depend on the fraction $\alpha
L_x$ of the system occupied by the stripe at the initial time (a single 
value $\alpha\approx1/3$ has been studied, yielding $\RE^{\perp}=195$).
In the same way, strong localised perturbations form spots
that make the whole system tumble into the turbulent state only for
$\RE>\RE^{\rm spots}\sim230$ but a precise determination of the
corresponding threshold is barely feasible since it implies a
three-parameter study by varying $\RE$, the initial size and amplitude
of the perturbation. Long range interactions associated to pressure
effects, well accounted for in the model, also seem to play an
important role.

To conclude, for decaying turbulence, the nucleation of sufficiently
large laminar domains seems to provide a good understanding of the origin
of the exponentially decreasing distribution of the transients' lifetimes
at given size, though a more complete study of the variation of the decay
rates with $\RE$ combined with size effects is needed, which is currently
underway. Though it has the
same observable consequence as the saddle interpretation, this new approach
clearly points to a spatiotemporal perspective that seems better suited
than the strictly temporal one since spatial extension is a crucial feature
of the problem. For onset, things are even much more complicated since
the transition depends sensitively on the shape and amplitude of the initial
condition and on the Reynolds number. Furthermore, it seems difficult to
find a definite connection between our different results obtained in a
spatiotemporal context and the search for {\it edge states\/} that have
been much studied recently in a finite dimensional framework
for both the pCf and the Ppf
\cite{Setal06,SEY07,Letal07}.

\section{Conclusion}

The transition to turbulence in wall flows leaves several problems open.
Most difficulties stem from the nature of the non-trivial solution
competing with the base state. Answers to this question have been
looked for first within linear stability theory extended
to take into account transient energy growth induced by non-normality,
and next using the theory of nonlinear dynamical systems and temporal
chaos. Accordingly, special periodic solutions (travelling waves)
have been found, that serve as a skeleton for complex dynamics described
in an abstract phase space in terms of {\it homoclinic tangles\/} and 
{\it chaotic transients\/}. This approach is however fully valid for
confined systems and can only be applied to open systems at the price of
putting artificial periodic boundary conditions at small distances in at
least one (Ppf is quasi-1D), if not two (pCf is quasi-2D) directions of
physical space. While the theory of chaotic transients well explains the
exponential behaviour of the lifetime distributions, it does not account
for the variation of their inverse decay rate with $\RE$, which was found
exponential in some cases and critical in others.

Taking another point of view, Pomeau long ago proposed an analogy of this
kind of discontinuous bifurcation to  a first-order (thermodynamic) phase
transition \cite{Po86} and put forward a related nucleation problem
\cite{Po98}. In the same time,
he introduced the concept of transition to turbulence {\it via
spatiotemporal intermittency\/}, a contamination process
where above some 
threshold, {\it activity\/} invade the system whereas it dies below it
\cite{Po86}.
The
spatio-temporally intermittent state is a mixture of active and quiescent
domains. Though at the time of the proposal, people focused mostly on
the universality properties of the continuous transition \cite{CM95},
examples of discontinuous transitions were known. One of them was put
in relation with the transition to turbulence in pCf \cite{BC98} but
the results were not reliable since the model was too far from
concrete hydrodynamics. Keeping all this in mind, we developed a model
which, instead of proceeding to full dimensional 
reduction in physical space, just froze part of the wall-normal dependence.
In spite of an insufficient energy transfer through cross-stream (small)
scales which led to a lowered transitional range, it correctly accounted
for the interplay of streamwise vortices and streaks (large in-plane
structures) and qualitatively reproduced hydrodynamical features, e.g.
non-local pressure effects, and transitional properties.
Even in the absence of firm conclusions (the work is still in progress),
the most interesting results are a better appreciation of the
drawbacks of the dynamical systems approach, and some support to the phase
transition viewpoint. It indeed suggests a different interpretation of
the transients' lifetime distribution in systems of moderate size and
offers a glimpse on the origin of complications arising from size effects
and the role of topology of laminar/turbulent domains in pCf. The
approach also suggests to look at Ppf along similar lines  by
considering it as a quasi-1D system and not as a 0D system in physical
space.

Apart from perspectives open for other wall flows and flow control, the
present approach and its spatiotemporal re-framing may help us better
understand the {\it nature\/} of the turbulent attractor
with respect to some ``thermodynamic'' approach to far-from-equilibrium
systems to be defined in a firm statistical physics environment.

\medskip

\noindent {\bf Acknowledgements:}\\
It is my pleasure to thank people at LadHyX: M.Lagha
(co-worker), C.Cossu, J.-M.Chomaz, P.Huerre, P.Schmid; at Saclay: S.Bottin,
O.Dauchot, F.Daviaud, A.Prigent; at Marburg: B.Eckhardt, J.Schumacher;
at ESPCI-Paris: L.Tuckerman, D.Barkley; at Bristol: R.R.Kerswell; at
ENS-Paris: Y.Pomeau.
All contributed at one level or another to my understanding of the
problem but, of course, none should be taken responsible for the views
expressed here. Large scale numerical simulations of the model were
performed within the framework of  projects \#6/1462 and \#6/2138 contracted
with IDRIS-Orsay.

\end{document}